# Strain tunable single-photon source based on a quantum dot-micropillar system

Magdalena Moczała-Dusanowska[1], Łukasz Dusanowski[1], Stefan Gerhardt[1], Yu Ming He[2], Marcus Reindl[3], Armando Rastelli[3], Rinaldo Trotta[3,4], Niels Gregersen[5], Sven Höfling[1,6], and Christian Schneider[1]

1 Technische Physik and Wilhelm Conrad Röntgen Research Center for Complex Material Systems, Physikalisches Institut, Würzburg University, Am Hubland, Würzburg, Germany
2 Hefei National Laboratory for Physical Sciences at Microscale and Department of Modern Physics, University of Science and Technology of China, Hefei, Anhui, 230026, China
3 Institute of Semiconductor and Solid State Physics, Johannes Kepler University, Altenbergerstr. 69, 4040 Linz, Austria
4 Department of Physics, Sapienza University of Rome, Piazzale Aldo Moro 5, 00185 Rome, Italy
5 DTU Fotonik, Department of Photonics Engineering, Technical University of Denmark, Ørsteds Plads 343, 2800 Kongens Lyngby, Denmark
6 SUPA, School of Physics and Astronomy, University of St Andrews, St Andrews, KY16 9SS, UK

ABSTRACT

Scalable quantum photonic architectures demand highly efficient, high-purity single-photon sources, which can be frequency matched via external tuning. We demonstrate a single-photon source based on an InAs quantum dot embedded in a micropillar resonator, which is frequency tunable via externally-applied stress. Our platform combines the advantages of a Bragg micropillar cavity and the piezo-strain-tuning technique enabling single photon spontaneous emission enhancement via the Purcell effect and quantum dot (QD) with tunable wavelength. Our optomechanical platform has been implemented by integration of semiconductor-based QD-micropillars on a piezoelectric substrate. The fabricated device exhibits spontaneous emission enhancement with a Purcell factor of 4.4±0.7 and allows for a pure triggered single-photon generation with $g^{(2)}(0) < 0.07$ under resonant excitation. A quantum dot emission energy tuning range of 0.75 meV for 27 kV/cm applied to the piezo substrate has been achieved. Our results pave the way towards the scalable implementation of single-photon quantum photonic technologies using optoelectronic devices.

Bright sources of indistinguishable single photons are key components for advanced quantum optics applications such as quantum communication in repeater architectures[1], boson sampling[2], and linear optical quantum computing[3]. Among different kinds of quantum emitters, semiconductor self-assembled quantum dots (QDs)[4] have been shown to be one of the prime candidates for the generation of single photons[5,6] as well as entangled photon pairs[7]. However, since QDs are embedded in a high refractive index material, light extraction in a homogeneous semiconductor medium is strongly limited due to total internal reflection. Photon extraction can be significantly enhanced by embedding QDs into photonic structures such as photonic crystal cavities[8], circular Bragg grating cavities[9], microlenses[10], nanowires[11], microdiscs[12] or micropillar cavities[13,14]. Among these architectures, QDs coupled to micropillar resonators have been shown to effectively utilize spontaneous emission enhancement via the Purcell effect[15], allowing to simultaneously achieve very high degrees of single photon indistinguishability and high photon extraction efficiencies. The narrowband spectral enhancement provided by the pillar cavities mitigates the detrimental effects of the phonon sideband emission on the indistinguishability[16,17] in contrast to broadband photonic structures. In combination with pulsed, resonant excitation schemes, single photon generation with simultaneous photon indistinguishability higher than 98%, purities above 99% and extraction efficiencies close of 66% were reported, thus outperforming any other solid state system [5,6,18].

The exploitation of the Purcell effect in high-Q cavities requires careful spectral alignment of the QD emission and the cavity resonance. This can be accomplished to some extent by deterministic fabrication of micropillar cavities[19,20], but usually a post-fabrication reversible spectral fine-tuning is still required. Spectral tuning is also critical for the implementation of identical multiple single-photon sources (SPSs), which can be operated within the spectral emission bandwidth (homogeneous broadening), and it seems unlikely that growing multiple QDs of exactly the same energy within the radiative linewidth will become possible. Thus, post-fabrication tuning of individual QD states to a common frequency seems to be the only realistic option.

For these reasons, several approaches have been developed in order to tune the emission energy of self-assembled QDs: Reversible tuning methods, which are feasible to adopt with a micropillar cavity device involve temperature tuning[20] as well as the application of external electric[21,22] and magnetic fields[23]. Spectral control via temperature is rather inadvisable for high-performance SPSs since the increase of temperature causes phonon-induced decoherence and consequently loss of photon indistinguishability[24,25]. On the other hand, electric-field-based QD energy control in micropillar cavities has been successfully demonstrated[18,21], however, electrical contacting is rather challenging due to the difficulties in fabricating reliable metal-semiconductor contacts compatible with resonant optical driving due to the undesired laser scattering from the metal surfaces.

Recently, strain tuning induced by integrated piezoelectric actuators has been successfully employed as a reliable and powerful tuning knob to control the optical properties of QDs[26–30]. Strain tuning allows for control of QD emission energy[31], polarization[32], exciton fine structure[27] or even light-hole admixture while maintaining high QD coherence[33]. Thus far, strain-tuning of QDs was mostly implemented in planar heterostructures[33,34] and broadband photonic structures providing no or only modest Purcell-enhancement to improve the photon extraction and indistinguishability. It was further utilized in different photonic structures, including nanowires[35,36], photonic crystal cavities[30,37,38], microlenses[39], and QDs integrated into photonic circuits[40], which were optimized for in-plane coupling. However, the implementation of a strain-tunable, resonant cavity platform, such as a micropillar, that directly utilizes the Purcell-enhancement for the implementation of a highly efficient single-photon source, has remained elusive.

In this work, we present the successful implementation of such a coupled QD-micropillar cavity platform integrated onto a piezoelectric actuator. By externally applied stress, we can tune the energy of QD transitions through the optical resonances of the microcavity, yielding a strong enhancement of spontaneous emission due to the Purcell effect and making it possible to characterize the system under various QD-cavity detuning.

DEVICE DESIGN AND FABRICATION

Our devices are based on epitaxially grown self-assembled In(Ga)As QDs, embedded in a planar microcavity. The epitaxial structures provide strong optical confinement in the growth direction by distributed Bragg reflectors (DBRs) composed of 15 and 25 periods of GaAs and AlAs layers in the top and bottom mirrors, respectively, surrounding a GaAs cavity.

The substrate was mechanically lapped down to a thickness of approximately 30 µm. Afterwards, an epoxy-based photoresist (SU8) was used to bond the sample to a 300 µm-thick [Pb(Mg$_{1/3}$Nb$_{2/3}$)O$_3$]$_{0.72}$[PbTiO$_3$]$_{0.28}$ (PMNPT) substrate side coated with chromium/gold contacts[27]. As a result, a device consisting of approximately 25 µm GaAs substrate and the planar microcavity with QDs attached to the piezoelectric actuator has been obtained. In the next steps, high-resolution electron beam lithography and a subsequent lift-off process were used to define the micropillars. The pattern was transferred to the sample using Reactive Ion Etching with an Ar/Cl$_2$ plasma. The etch depth of the micropillars was carefully adjusted, such that only the top DBR mirrors, the cavity region and finally two to four bottom DBR mirror pairs were etched. In Figure 1(a) we depict an artistic sketch of the device. A scanning electron microscope (SEM) image showing an exemplary etched micropillar with a diameter *d* of about 2.5 µm and a height of about 2.6 µm is presented in the inset of Fig. 1(b). As the last step, planarization using benzocyclobutene (BCB) polymer was performed to mechanically stabilize the pillars and protect the etched sidewalls from oxidation. The final devices were mounted onto an AlN chip carrier, providing heat transfer and electrical contacts to the piezoelectric actuator via wire bonding.

RESULTS AND DISCUSSION

Our micropillar devices were characterized by means of micro-photoluminescence (µPL) spectroscopy at a nominal temperature of around 4.5 K. For all experiments, the QDs were excited optically through a microscope objective with a numerical aperture (NA) of 0.4. The signal emitted from the sample was collected by the same objective, then introduced into a single mode optical fiber and finally analyzed by a 0.75 m focal length monochromator equipped with a liquid-nitrogen-cooled silicon CCD detector. For resonance fluorescence experiments, the cross-polarization configuration was used and the QDs were excited by a tunable 2 ps pulsed laser with 82 MHz repetition rate. For time-resolved experiments, we utilized a fast avalanche photo-diode (APD) with 40 ps time-resolution. Second-order photon correlation measurements were performed in a Hanbury Brown and Twiss configuration by utilizing a fiber-based beam splitter and a pair of single photon counting APDs with 350 ps resolution. The photon correlation and time-resolved emission events were acquired by a multichannel picosecond event timer.

The µPL spectrum recorded under non-resonant cw excitation (532 nm laser line) for device #1 with a 2.5 µm-diameter micropillar is shown in Fig. 1(b). Several sharp peaks stemming from the emission from QDs and a wider peak which can be assigned to the fundamental cavity mode are observed. The optical cavity mode peak centered 1.365 eV energy is characterized by a full-width at half-maximum (FWHM) of 327 µeV, corresponding to a quality factor Q of 3660. We note that for micropillars with similar diameters, which were etched through the whole bottom DBR mirror, we would expect an increase of the maximally obtainable Q factors by approximately a factor of two. However, etching the full structure would strongly limit the amount of strain transferred from the substrate to the QDs. To study how the etching depth influences the Purcell factor obtainable in our micropillars, we have performed numerical simulations using a Fourier Modal Method[41] of the Purcell factor as a function of the number of etched bottom DBR layer pairs and pillar diameter. Indeed, the reduction of the quality factor, in combination with increased mode volumes, yield a reduction of the maximally achievable Purcell factors, as displayed in Fig. 1(c). A pillar diameter in the range of 2-3 µm[42,43] is preferred to ensure a narrow beam divergence and a high collection efficiency. In this diameter range, a good compromise is obtained for a 2-4 of etched DBR layer pairs for which the Purcell factor is only slightly reduced.

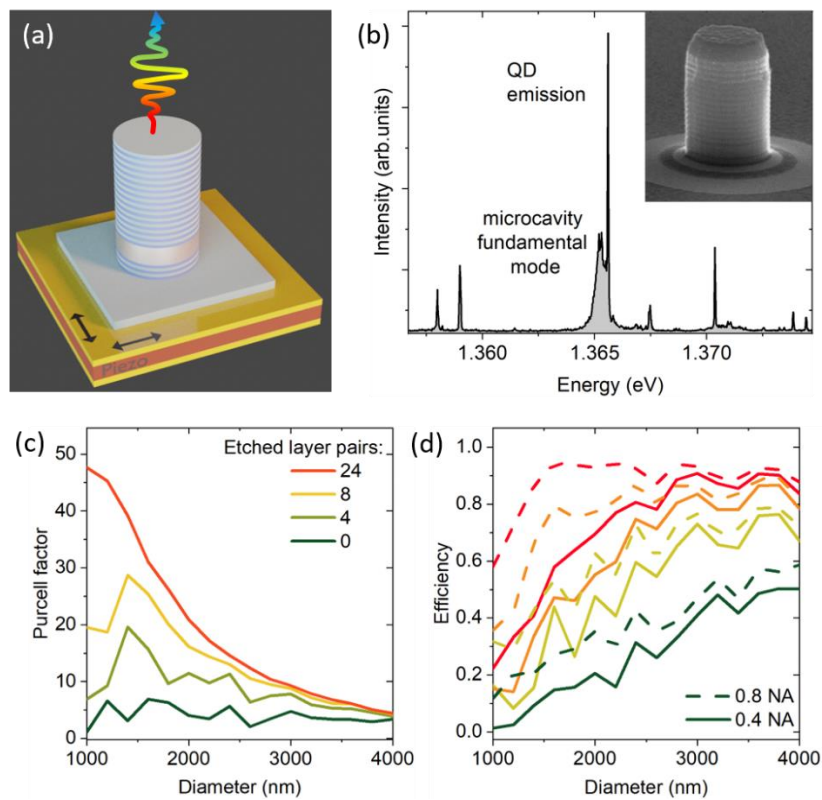

FIG. 1. (a) Artistic sketch of a QD-micropillar cavity integrated with the piezoelectric substrate. (b) Micro-photoluminescence spectrum of a micropillar cavity with embedded QDs (at a temperature of 4.5 K). Visible are the microcavity fundamental mode and QD emission lines. The microcavity exhibits a quality factor of around Q = 3660. Inset: SEM image of the studied micropillar cavity (before planarization) with a diameter of d=2.5 µm. (c) Numerical simulations of the Purcell factor as a function of pillar diameter *d* and numbers of bottom DBR etched layer pairs. (d) Numerical simulations of the extraction efficiency as a function of pillar diameter d and numbers of bottom DBR etched layer pairs for 0.4 NA and 0.8 NA collection optics.

Numerical simulations of the collection efficiency of our devices, assuming a numerical aperture of 0.4 and 0.8, as a function of the number of etched bottom DBR layer pairs and pillar diameter have been performed using a Fourier modal method[41] and are presented in Fig. 1(d). For shallow etched micropillars, the efficiency as well as the Purcell factor are slightly reduced by around 10-40%, in comparison to a fully etched micropillar resonator, depending on the number of the etched DBR mirrors. For the case of four etched DBR layers underneath the cavity, one can expect extraction efficiencies up to 73% (77%) for micropillars with dimeters of 2-3 µm and 0.4 (0.8) NA collection optics. Figure 2 depicts the QD emission tunability of the device #1 as a function of the voltage applied to the PMNPT. The PMNPT was poled such that a positive (negative) voltage leads to in-plane compression (expansion) of the substrate and thus of the micropillars. Figure 2(a) shows a color-coded µPL spectral intensity map (logarithmic scale) recorded for QDs located in device #1 as a function of the voltage applied to the piezo substrate. The applied voltage was swept in the range from −350 V to 450 V, ramped in steps of 5 V. As expected[34], the compressive (tensile) biaxial strain results in an increase (decrease) of the QD emission energy. A statistical analysis carried out on more than 100 QDs yielded, that the emission energy of the majority of the QDs placed within the same micropillar show very similar tuning range and the ratio (within a 15 % margin). On the other hand, the cavity mode is rather insensitive to the applied strain, and shift only 30 µeV. This particular feature, which has been already observed and explained for QD in planar cavities[28], gives us the possibility to design a cavity for a specific wavelength and use strain to tune the QD emission through the cavity mode. In Figure 2(b), we display a zoom on one QD emission line in the close spectral vicinity of the microcavity mode (plotted in linear color scale). As we apply external stress to our system, we can spectrally tune this QD through the cavity mode and observe a strong enhancement of the QD emission at spectral resonance with the cavity mode.

The relation between the QD emission energy and the applied electric field to piezo is plotted in Fig. 2(c). We extract an overall tuning range of 745 µeV (0.49 nm) over an electric field modification of 27 kV/cm. A linear relationship between these two parameters is clearly observed with a strain-tuning slope of 0.93 µeV/V. We note that there exists a trade-off between the Q factor of the cavity and the obtainable tuning range. The tuning range of our device is increased by reducing the number of etched mirror pairs in the bottom DBR at the price of a reduction of the obtainable Q factor and thus, the Purcell factor. For deeply etched devices, we anticipate that strain relaxation in the bottom DBR section limits the obtainable tuning range. Therefore, etching a few bottom DBR pairs seems to be the optimal choice for strain-tunability of QD emission lines and Q factor, as discussed in more detail below.

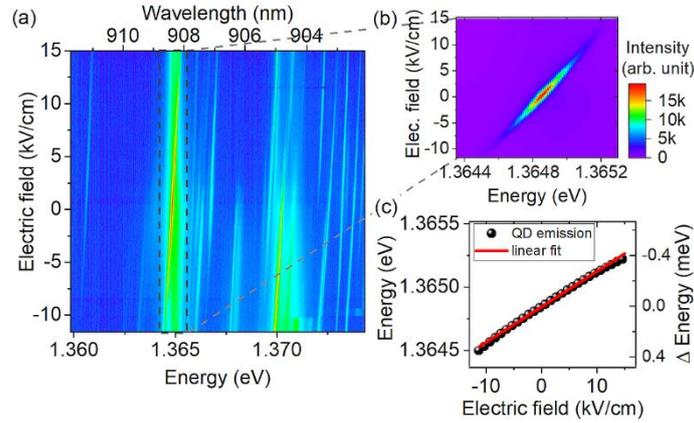

FIG. 2. (a and b) Color-coded µPL intensity spectral maps of the QDs emission as a function of the electric field applied to the piezo substrate. (a) Broadband µPL emission map of several QDs located in one micropillar cavity (logarithmic intensity scale). Most of the emission lines follow the same spectral shift. (b) Zoom into the µPL intensity map in the area of the fundamental cavity mode (linear intensity scale). A strong enhancement of the emission due to the Purcell effect at spectral resonance is observed. (c) QD emission energy as a function of applied piezo-voltage. The application of external stress via the piezocrystal induces a linear shift of the QD emission energy up to 0.75 meV (0.93 µeV/V).

For further studies, we investigated a similar device (micropillar #2) with a diameter of $d$ = 2.8 µm with four etched bottom DBR pairs and a slightly improved cavity quality factor of Q = 4388. For this device, a theoretical analysis shows that the attainable value of the Purcell factor is around 7. In this case, the QD emission at 4.5 K is slightly red detuned from the fundamental cavity mode resonance (180 µeV) and thus cannot be brought into resonance by increasing the temperature. Yet, by applying a positive piezo-voltage to our system, it is possible to tune the QD almost completely into the cavity resonance as depicted in Fig. 3(a).

In the inset of Fig. 3(a), the color-coded µPL intensity map of the QD emission as a function of applied voltage is shown. As the applied positive piezo-voltage is increased, the QD lines blue-shift (120 µeV for 13.4 kV/cm), and can be tuned into resonance with the cavity mode. At the same time, we observe a significant increase of the QD emission intensity, indicating the spontaneous emission enhancement due to the Purcell effect.

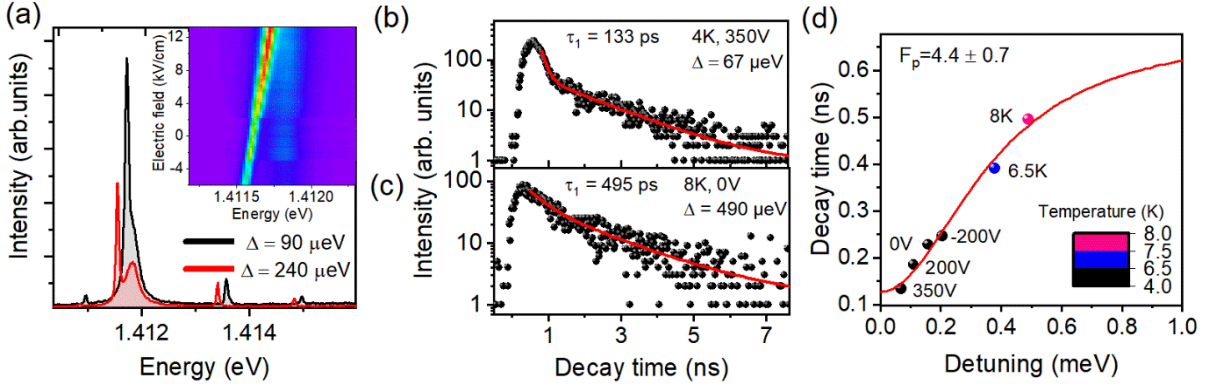

FIG. 3. (a) Micro-photoluminescence spectrum of a micropillar #2 for positive and negative applied voltages resulting in different energy detuning Δ of QD emission from the cavity mode. By applying a positive piezo-voltage, the QD emission can be tuned into resonance with the microcavity fundamental mode. Inset: Color-coded µPL intensity map of the QD emission as a function of the voltage applied to the piezo substrate. A strong emission enhancement at spectral resonance due to the Purcell effect is observed. (b) Time-resolved resonance fluorescence measurements of the strain-tunable single-photon source. Resonance fluorescence time-traces of the QD emission almost at spectral resonance with the fundamental cavity mode (67 µeV detuning) and (c) detuned by 490 µeV. Red lines indicate the double exponential fit, where a fast decay corresponds to the QD emission into the micropillar fundamental mode. (d) Resonance fluorescence decay time (fast component) as a function of QD-cavity-resonance detuning. Black points are recorded at 4.5 K temperature and various piezo-voltages; color points are recorded under varied temperatures. The red solid line is a fit to the QD lifetime vs QD-cavity detuning. A maximum Purcell factor of 4.4±0.7 has been extracted from a fitting procedure.

To estimate the Purcell enhancement factor for a device #2, we performed time-resolved resonance fluorescence measurements. For that purpose, the QD has been gradually tuned into resonance and out of resonance with the fundamental cavity mode using a combination of piezo- and temperature-tuning. A PL decay curve recorded for the QD emission almost perfectly tuned into cavity mode resonance (67 µeV detuning) is presented in Fig. 3(b). The recorded emission time-trace exhibits a two-component exponential decay. We associate the fast decay with the Purcell-enhanced spontaneous emission of the QD state into the fundamental cavity mode, while the slower component is interpreted as the cavity mode emission fed by other QD transitions, which are detuned from the mode resonance[44,45]. Our experimental data have been fitted using a double exponential function indicated in the graph by a red solid line. Upon the fitting procedure, the characteristic decay time $\tau_1$ = 133 ps is extracted for the fast component. Figure 3(c) shows the time-resolved resonance fluorescence measurement for the QD detuned from the cavity mode by 490 µeV. As in the previous case, we observe the double exponential decay, with the faster decay time equal to around 495 ps. This observation proves that our QD-micropillar device indeed exhibits Purcell enhancement. To quantify more precisely the Purcell enhancement, the decay time traces have been recorded for several different voltages applied to the piezo substrate at a temperature of 4.5 K and two additional measurements performed at temperatures of 6.5 K and 8 K (to enable slightly larger detunings). The extracted characteristic emission times $\tau_1$ were then plotted as a function of the emitter-cavity detuning in Fig. 3(d).

This analysis enables us to accurately determine the Purcell factor of our QD-cavity device by fitting the decay time $\tau$ as a function of the detuning $\Delta$ via[46]

$$\tau(\Delta) = \frac{F_{P(\max)}}{F_{P(\max)}\delta + 1} \cdot \frac{\hbar \epsilon_0 V_m}{2Q\mu_{12}^2}$$

where $F_{P(\max)} = \frac{3Q(\lambda/n)^3}{4\pi^2 V_m}$ is the maximal Purcell factor, $\delta = \xi^2 \cdot \frac{\Delta\omega_c^2}{4(\Delta^2)+\Delta\omega_c^2}$, $\xi$ represents the orientation mismatch between the local cavity field and the dipole moment of the QD, and $\Delta\omega_c$ is the linewidth of the cavity mode. $\mu_{12}$ is the dipole moment of the radiative transition and Q and $V_m$ are the quality factor and the mode volume of the cavity mode, respectively. The analysis yields a Purcell factor of 4.4±0.7, which is in qualitative agreement with the Purcell-factor derived for a shallow etched pillar microcavity as it is depicted in Fig. 1(c). In order to estimate the brightness of our experimentally investigated micropillar device, we calculated the beta factor $\beta$, which describes the emitter coupling efficiency into the pillar cavity mode. Following the formula $\beta = F_P/(F_P + 1)$ for Purcell factor $F_P$ equal to 4.4±0.7, we estimated that $\beta \approx 80\%$.

In Figure 4(a), we plot the power dependency of our pulsed resonance fluorescence signal, measured for QD-cavity detuning of $\Delta$ = 180 µeV. A clear oscillatory behavior is observed, which is a hallmark of the damped Rabi oscillations of a coherently driven two-level system[47]. The resonance fluorescence intensity reaches a maximum for the laser power of 223 nW, corresponding to a π–pulse, where the system reaches inversion. The red solid line is a fit using a damped Rabi rotation model[25]. Within this approximation, the dominant phonon coupling effects are captured by a temperature-dependent Rabi frequency renormalization.

Under these conditions, we confirm the single-photon character of the generated resonance fluorescence photons in a second-order correlation experiment. In Figure 4(b), we plot the normalized second-order correlation function $g^{(2)}(\tau)$ histogram recorded for the considered QD transition as a function of delay time between subsequently emitted photons. The central peak is strongly suppressed, a core signature of a single photon source. In order to derive the purity of the source, we divide the integrated counts of the central peak by the mean value of the adjacent set of peaks, yielding a $g^{(2)}(0)$ value of 0.07±0.02.

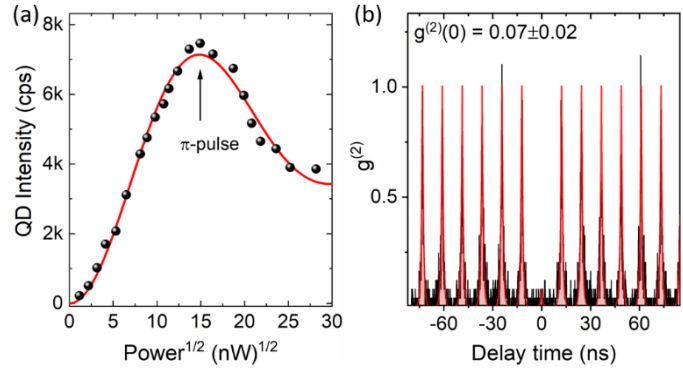

FIG. 4. (a) Resonance fluorescence intensity plotted versus the square root of the excitation power, recorded for QD-cavity detuning of $\Delta$ = 180 µeV. The red solid line is a fit using a damped Rabi rotation model. (b) Second-order autocorrelation histogram as a function of delay time between subsequently emitted photons recorded under pulsed s-shell resonant excitation with a π-pulse area (223 nW). The red solid line is a two-sided exponential decay fit.

In summary, we have realized a strain-tunable single-photon source by integrating a QD-pillar microcavity onto a piezoelectric substrate. We apply external stress to induce a linear shift of the QD emission energy (0.75 meV for 27 kV/cm), which is sufficiently large to tune QDs through the spectral resonances of our pillar cavity. We used a partially etched pillar geometry that allows achieving an optimal trade-off between spontaneous emission enhancement and the transfer of strain from the piezoelectric substrate. Using this approach, we demonstrated that it is possible to tune the QD emission energy in resonance with the cavity mode and obtain an appreciable Purcell enhancement factor of around 4.4, despite the quality factor being limited to ~4500. Further development of the tuning capabilities could be realized following one of two approaches: (i) application of more advanced planarization schemes[40] to increase strain transfer from substrate to layer containing QDs, (ii) using more powerful piezo substrates such as a micromachined ones[33,48], where much higher strain could be potentially achieved, and thus lead into increase of spectral tuning range. Under resonant pulsed excitation conditions, we demonstrated that our device maintains high purity single-photon generation with $g^{(2)}(0) = 0.07\pm0.02$. The presented platform makes it possible to fully exploit the advantages of the micropillar structure, with extraction efficiencies up to 73% (77%) for micropillars with dimeters of 2-3 µm. Due to the applied resonant excitation scheme, the time uncertainty of the exciton creation has been most likely eliminated[6,18] so that our source may be useful for linear optical quantum computing protocols, where a high degree of photon indistinguishability is demanded. Spectral tuning capabilities of our single photon source pave the way towards the scalable quantum photonic architectures, where multiple single photon sources need to be engineered and tuned into a common energy resonance.

ACKNOWLEDGMENTS

The authors acknowledge Silke Kuhn's technical support. Ł.D. acknowledges financial support from the Alexander-von-Humboldt Foundation. This work has been supported by the State of Bavaria and the QuantERA HYPER-U-P-S project. Project HYPER-U-P-S has received funding from the QuantERA ERA-NET Cofund in Quantum Technologies implemented within the European Union's Horizon 2020 Programme.

REFERENCES

(1) Sangouard, N.; Simon, C.; Minář, J.; Zbinden, H.; de Riedmatten, H.; Gisin, N. Long-Distance Entanglement Distribution with Single-Photon Sources. *Phys. Rev. A* **2007**, *76*, 050301.

(2) Wang, H.; He, Y.-M.; Li, Y.-H.; Su, Z.-E.; Li, B.; Huang, H.-L.; Ding, X.; Chen, M.-C.; Liu, C.; Qin, J.; et al. High-Efficiency Multiphoton Boson Sampling. *Nat. Photonics* **2017**, *11*, 361–365.

(3) Carolan, J.; Harrold, C.; Sparrow, C.; Martin-Lopez, E.; Russell, N. J.; Silverstone, J. W.; Shadbolt, P. J.; Matsuda, N.; Oguma, M.; Itoh, M.; et al. Universal Linear Optics. *Science (80-. ).* **2015**, *349*, 711–716.

(4) Marzin, J.-Y.; Gérard, J.-M.; Izraël, A.; Barrier, D.; Bastard, G. Photoluminescence of Single InAs Quantum Dots Obtained by Self-Organized Growth on GaAs. *Phys. Rev. Lett.* **1994**, *73*, 716–719.

(5) Unsleber, S.; He, Y.-M.; Gerhardt, S.; Maier, S.; Lu, C.-Y.; Pan, J.-W.; Gregersen, N.; Kamp, M.; Schneider, C.; Höfling, S. Highly Indistinguishable On-Demand Resonance Fluorescence Photons from a Deterministic Quantum Dot Micropillar Device with 74% Extraction Efficiency. *Opt. Express* **2016**, *24*, 8539.

(6) Ding, X.; He, Y.; Duan, Z.-C.; Gregersen, N.; Chen, M.-C.; Unsleber, S.; Maier, S.; Schneider, C.; Kamp, M.; Höfling, S.; et al. On-Demand Single Photons with High Extraction Efficiency and Near-Unity Indistinguishability from a Resonantly Driven Quantum Dot in a Micropillar. *Phys. Rev. Lett.* **2016**, *116*, 020401.

(7) Huber, D.; Reindl, M.; Covre da Silva, S. F.; Schimpf, C.; Martín-Sánchez, J.; Huang, H.; Piredda, G.; Edlinger, J.; Rastelli, A.; Trotta, R. Strain-Tunable GaAs Quantum Dot: A Nearly Dephasing-Free Source of Entangled Photon Pairs on Demand. *Phys. Rev. Lett.* **2018**, *121*, 033902.

(8) Englund, D.; Fattal, D.; Waks, E.; Solomon, G.; Zhang, B.; Nakaoka, T.; Arakawa, Y.; Yamamoto, Y.; Vučković, J. Controlling the Spontaneous Emission Rate of Single Quantum Dots in a Two-Dimensional Photonic Crystal. *Phys. Rev. Lett.* **2005**, *95*, 013904.

(9) Davanço, M.; Rakher, M. T.; Schuh, D.; Badolato, A.; Srinivasan, K. A Circular Dielectric Grating for Vertical Extraction of Single Quantum Dot Emission. *Appl. Phys. Lett.* **2011**, *99*, 1–4.

(10) Gschrey, M.; Thoma, A.; Schnauber, P.; Seifried, M.; Schmidt, R.; Wohlfeil, B.; Krüger, L.; Schulze, J. H.; Heindel, T.; Burger, S.; et al. Highly Indistinguishable Photons from Deterministic Quantum-Dot Microlenses Utilizing Three-Dimensional in Situ Electron-Beam Lithography. *Nat. Commun.* **2015**, *6*, 7662.

(11) Friedler, I.; Sauvan, C.; Hugonin, J. P.; Lalanne, P.; Claudon, J.; Gérard, J. M. Solid-State Single Photon Sources: The Nanowire Antenna. *Opt. Express* **2009**, *17*, 2095.

(12) Balram, K. C.; Davanço, M.; Lim, J. Y.; Song, J. D.; Srinivasan, K. Moving Boundary and Photoelastic Coupling in GaAs Optomechanical Resonators. *Optica* **2014**, *1*, 414.

(13) Moreau, E.; Robert, I.; Gérard, J. M.; Abram, I.; Manin, L.; Thierry-Mieg, V. Single-Mode Solid-State Single Photon Source Based on Isolated Quantum Dots in Pillar Microcavities. *Appl. Phys. Lett.* **2001**, *79*, 2865–2867.

(14) Pelton, M.; Santori, C.; Vučković, J.; Zhang, B.; Solomon, G. S.; Plant, J.; Yamamoto, Y. Efficient Source of Single Photons: A Single Quantum Dot in a Micropost Microcavity.

*Phys. Rev. Lett.* **2002**, *89*, 233602.

(15) Gérard, J.; Sermage, B.; Gayral, B.; Legrand, B.; Costard, E.; Thierry-Mieg, V. Enhanced Spontaneous Emission by Quantum Boxes in a Monolithic Optical Microcavity. *Phys. Rev. Lett.* **1998**, *81*, 1110–1113.

(16) Grange, T.; Somaschi, N.; Antón, C.; De Santis, L.; Coppola, G.; Giesz, V.; Lemaître, A.; Sagnes, I.; Auffèves, A.; Senellart, P. Reducing Phonon-Induced Decoherence in Solid-State Single-Photon Sources with Cavity Quantum Electrodynamics. *Phys. Rev. Lett.* **2017**, *118*, 253602.

(17) Iles-Smith, J.; McCutcheon, D. P. S.; Nazir, A.; Mørk, J. Phonon Scattering Inhibits Simultaneous Near-Unity Efficiency and Indistinguishability in Semiconductor Single-Photon Sources. *Nat. Photonics* **2017**, *11*, 521–526.

(18) Somaschi, N.; Giesz, V.; De Santis, L.; Loredo, J. C.; Almeida, M. P.; Hornecker, G.; Portalupi, S. L.; Grange, T.; Antón, C.; Demory, J.; et al. Near-Optimal Single-Photon Sources in the Solid State. *Nat. Photonics* **2016**, *10*, 340–345.

(19) Dousse, A.; Lanco, L.; Suffczyński, J.; Semenova, E.; Miard, A.; Lemaître, A.; Sagnes, I.; Roblin, C.; Bloch, J.; Senellart, P. Controlled Light-Matter Coupling for a Single Quantum Dot Embedded in a Pillar Microcavity Using Far-Field Optical Lithography. *Phys. Rev. Lett.* **2008**, *101*, 30–33.

(20) He, Y.-M.; Liu, J.; Maier, S.; Emmerling, M.; Gerhardt, S.; Davanço, M.; Srinivasan, K.; Schneider, C.; Höfling, S. Deterministic Implementation of a Bright, on-Demand Single-Photon Source with near-Unity Indistinguishability via Quantum Dot Imaging. *Optica* **2017**, *4*, 802.

(21) Kistner, C.; Heindel, T.; Schneider, C.; Rahimi-Iman, A.; Reitzenstein, S.; Höfling, S.; Forchel, A. Demonstration of Strong Coupling via Electro-Optical Tuning in High-Quality QD-Micropillar Systems. *Opt. Express* **2008**, *16*, 15006.

(22) Nowak, A. K.; Portalupi, S. L.; Giesz, V.; Gazzano, O.; Dal Savio, C.; Braun, P.-F.; Karrai, K.; Arnold, C.; Lanco, L.; Sagnes, I.; et al. Deterministic and Electrically Tunable Bright Single-Photon Source. *Nat. Commun.* **2014**, *5*, 3240.

(23) Reitzenstein, S.; Münch, S.; Franeck, P.; Rahimi-Iman, A.; Löffler, A.; Höfling, S.; Worschech, L.; Forchel, A. Control of the Strong Light-Matter Interaction between an Elongated $In_{0.3}Ga_{0.7}As$ Quantum Dot and a Micropillar Cavity Using External Magnetic Fi. *Phys. Rev. Lett.* **2009**, *103*, 127401.

(24) Thoma, A.; Schnauber, P.; Gschrey, M.; Seifried, M.; Wolters, J.; Schulze, J.-H.; Strittmatter, A.; Rodt, S.; Carmele, A.; Knorr, A.; et al. Exploring Dephasing of a Solid-State Quantum Emitter via Time- and Temperature-Dependent Hong-Ou-Mandel Experiments. *Phys. Rev. Lett.* **2016**, *116*, 033601.

(25) Gerhardt, S.; Iles-Smith, J.; McCutcheon, D. P. S.; He, Y.-M.; Unsleber, S.; Betzold, S.; Gregersen, N.; Mørk, J.; Höfling, S.; Schneider, C. Intrinsic and Environmental Effects on the Interference Properties of a High-Performance Quantum Dot Single-Photon Source. *Phys. Rev. B* **2018**, *97*, 195432.

(26) Zander, T.; Herklotz, A.; Kiravittaya, S.; Benyoucef, M.; Ding, F.; Atkinson, P.; Kumar, S.; Plumhof, J. D.; Dörr, K.; Rastelli, A.; et al. Epitaxial Quantum Dots in Stretchable Optical Microcavities. *Opt. Express* **2009**, *17*, 22452–22461.

(27) Trotta, R.; Martín-Sánchez, J.; Wildmann, J. S.; Piredda, G.; Reindl, M.; Schimpf, C.; Zallo, E.; Stroj, S.; Edlinger, J.; Rastelli, A. Wavelength-Tunable Sources of Entangled Photons Interfaced with Atomic Vapours. *Nat. Commun.* **2016**, *7*, 10375.

(28) Trotta, R.; Atkinson, P.; Plumhof, J. D.; Zallo, E.; Rezaev, R. O.; Kumar, S.; Baunack, S.; Schröter, J. R.; Rastelli, A.; Schmidt, O. G. Nanomembrane Quantum-Light-Emitting Diodes Integrated onto Piezoelectric Actuators. *Adv. Mater.* **2012**, *24*, 2668–2672.

(29) Trotta, R.; Rastelli, A. Engineering of Quantum Dot Photon Sources via Electro-Elastic Fields. In: Pre. In *Engineering the Atom-Photon Interaction. Nano-Optics and Nanophotonics.*; Predojević, A., Mitchell, M., Eds.; Springer, Cham, 2015.

(30) Sun, S.; Kim, H.; Solomon, G. S.; Waks, E. Strain Tuning of a Quantum Dot Strongly Coupled to a Photonic Crystal Cavity. *Appl. Phys. Lett.* **2013**, *103*, 151102.

(31) Kumar, S.; Trotta, R.; Zallo, E.; Plumhof, J. D.; Atkinson, P.; Rastelli, A.; Schmidt, O. G. Strain-Induced Tuning of the Emission Wavelength of High Quality GaAs/AlGaAs Quantum Dots in the Spectral Range of the $^{87}$Rb D$_2$ Lines. *Appl. Phys. Lett.* **2011**, *99*, 161118.

(32) Plumhof, J. D.; Trotta, R.; Křápek, V.; Zallo, E.; Atkinson, P.; Kumar, S.; Rastelli, A.; Schmidt, O. G. Tuning of the Valence Band Mixing of Excitons Confined in GaAs/AlGaAs Quantum Dots via Piezoelectric-Induced Anisotropic Strain. *Phys. Rev. B - Condens. Matter Mater. Phys.* **2013**, *87*, 33–35.

(33) Yuan, X.; Weyhausen-Brinkmann, F.; Martín-Sánchez, J.; Piredda, G.; Křápek, V.; Huo, Y.; Huang, H.; Schimpf, C.; Schmidt, O. G.; Edlinger, J.; et al. Uniaxial Stress Flips the Natural Quantization Axis of a Quantum Dot for Integrated Quantum Photonics. *Nat. Commun.* **2018**, *9*, 1–8.

(34) Ding, F.; Singh, R.; Plumhof, J. D.; Zander, T.; Křápek, V.; Chen, Y. H.; Benyoucef, M.; Zwiller, V.; Dörr, K.; Bester, G.; et al. Tuning the Exciton Binding Energies in Single Self-Assembled InGaAs/GaAs Quantum Dots by Piezoelectric-Induced Biaxial Stress. *Phys. Rev. Lett.* **2010**, *104*, 2–5.

(35) Kremer, P. E.; Dada, A. C.; Kumar, P.; Ma, Y.; Kumar, S.; Clarke, E.; Gerardot, B. D. Strain-Tunable Quantum Dot Embedded in a Nanowire Antenna. *Phys. Rev. B - Condens. Matter Mater. Phys.* **2014**, *90*, 201408.

(36) Yeo, I.; De Assis, P. L.; Gloppe, A.; Dupont-Ferrier, E.; Verlot, P.; Malik, N. S.; Dupuy, E.; Claudon, J.; Gérard, J. M.; Auffèves, A.; et al. Strain-Mediated Coupling in a Quantum Dot-Mechanical Oscillator Hybrid System. *Nat. Nanotechnol.* **2014**, *9*, 106–110.

(37) Beetz, J.; Braun, T.; Schneider, C.; Höfling, S.; Kamp, M. Anisotropic Strain-Tuning of Quantum Dots inside a Photonic Crystal Cavity. *Semicond. Sci. Technol.* **2013**, *28*.

(38) Luxmoore, I. J.; Ahmadi, E. D.; Luxmoore, B. J.; Wasley, N. A.; Tartakovskii, A. I.; Hugues, M.; Skolnick, M. S.; Fox, A. M. Restoring Mode Degeneracy in H1 Photonic Crystal Cavities by Uniaxial Strain Tuning. *Appl. Phys. Lett.* **2012**, *100*, 121116.

(39) Fischbach, S.; Helversen, M. v.; Schmidt, M.; Kaganskiy, A.; Schmidt, R.; Schliwa, A.; Heindel, T.; Rodt, S.; Reitzenstein, S. A Deterministically Fabricated Spectrally-Tunable Quantum Dot Based Single-Photon Source. **2018**.

(40) Elshaari, A. W.; Büyüközer, E.; Zadeh, I. E.; Lettner, T.; Zhao, P.; Schöll, E.; Gyger, S.; Reimer, M. E.; Dalacu, D.; Poole, P. J.; et al. Strain-Tunable Quantum Integrated Photonics. *Nano Lett.* **2018**, acs.nanolett.8b03937.

(41) Häyrynen, T.; de Lasson, J. R.; Gregersen, N. Open-Geometry Fourier Modal Method: Modeling Nanophotonic Structures in Infinite Domains. *J. Opt. Soc. Am. A* **2016**, *33*, 1298.

(42) Barnes, W. L.; Björk, G.; Gérard, J. M.; Jonsson, P.; Wasey, J. A. E.; Worthing, P. T.; Zwiller, V. Solid-State Single Photon Sources: Light Collection Strategies. *Eur. Phys. J. D - At. Mol. Opt. Phys.* **2002**, *18*, 197–210.

(43) Reitzenstein, S.; Forchel, A. Quantum Dot Micropillars. *J. Phys. D. Appl. Phys.* **2010**, *43*, 033001.

(44) Hennessy, K.; Badolato, A.; Winger, M.; Gerace, D.; Atatüre, M.; Gulde, S.; Fält, S.; Hu,

E. L.; Imamoğlu, A. Quantum Nature of a Strongly Coupled Single Quantum Dot–cavity System. *Nature* **2007**, *445*, 896–899.

(45) Suffczyński, J.; Dousse, A.; Gauthron, K.; Lemaître, A.; Sagnes, I.; Lanco, L.; Bloch, J.; Voisin, P.; Senellart, P. Origin of the Optical Emission within the Cavity Mode of Coupled Quantum Dot-Cavity Systems. *Phys. Rev. Lett.* **2009**, *103*, 027401.

(46) He, Y.-M.; Liu, J.; Maier, S.; Emmerling, M.; Gerhardt, S.; Davanço, M.; Srinivasan, K.; Schneider, C.; Höfling, S. Deterministic Implementation of a Bright, on-Demand Single-Photon Source with near-Unity Indistinguishability via Quantum Dot Imaging. *Optica* **2017**, *4*, 802.

(47) Muller, A.; Flagg, E. B.; Bianucci, P.; Wang, X. Y.; Deppe, D. G.; Ma, W.; Zhang, J.; Salamo, G. J.; Xiao, M.; Shih, C. K. Resonance Fluorescence from a Coherently Driven Semiconductor Quantum Dot in a Cavity. *Phys. Rev. Lett.* **2007**, *99*, 187402.

(48) Martín-Sánchez, J.; Trotta, R.; Piredda, G.; Schimpf, C.; Trevisi, G.; Seravalli, L.; Frigeri, P.; Stroj, S.; Lettner, T.; Reindl, M.; et al. Reversible Control of In-Plane Elastic Stress Tensor in Nanomembranes. *Adv. Opt. Mater.* **2016**, *4*, 682–687.